# Magnetic susceptibility and heat capacity of a novel antiferromagnet: LiNi$_2$P$_3$O$_{10}$ and the effect of doping

P Khuntia and A V Mahajan

*Department of Physics, Indian Institute of Technology Bombay, Powai, Mumbai-400076, India*

We report the synthesis, x-ray diffraction, magnetic susceptibility and specific heat measurements on polycrystalline samples of undoped LiNi$_2$P$_3$O$_{10}$ and samples with non-magnetic impurity (Zn$^{2+}$, $S = 0$) and magnetic impurity (Cu$^{2+}$, $S = 1/2$) at the Ni site. The magnetic susceptibility data show a broad maximum at around 10 K and a small anomaly at about 7 K in the undoped sample. There is a λ-like anomaly in the specific heat at 7 K , possibly due to the onset of antiferromagnetic ordering in the system. The magnetic entropy change at the ordering temperature is close to the value corresponding to $R$ln (2$S$+1) expected for an $S = 1$ system. The temperature corresponding to the broad maximum and the ordering temperature both decrease on Zn and Cu substitutions and also in applied magnetic fields.

## I. INTRODUCTION

The magnetic properties of pure and doped low dimensional quantum spin systems have been of interest to the condensed matter community following the discovery of many exotic and interesting phenomena in these systems. Especially in cases with a low value of the spin such as $S = 1/2$ there are strong quantum fluctuations and novel ground states can result. In some cases simple magnetic models can be used and thus, theoretical and numerical simulations can directly be compared with experimental results. Low dimensionality and strong electron correlations have been found to play a key role in driving novel phenomena like quantum criticality, spin–glass transition, spin liquids, valence bond solid, superconductivity and many more exotic ground states [1-4].

Generally, ideal one dimensional (1D) Heisenberg antiferromagnets do not exhibit long range ordering at low temperature due to strong quantum fluctuations. However in real materials, weak interchain interactions are responsible for the three dimensional (3D) ordering [5]. Application of high magnetic field can suppress the interchain interaction leading the system from 3D ordered state to 1D quantum disordered state [6]. Considerable amount of attention has been paid to the study of the phase diagram of weakly anisotropic antiferromagnets where the transition to a long range order has been induced by Ising type anisotropy or interchain or interlayer coupling [7,8]. Theoretically, Néel [9] and experimentally Poulis *et al*. [10] have shown that, the applied magnetic field along the easy axis of an antiferromagnet(AFM) can rotate the sublattice vectors by $90^o$ to a spin–flop (SF) phase and the AFM to SF transition is a first order phase transition associated with a sudden rise of magnetisation with increasing applied magnetic field. Recently, much attention has been paid to Cu and V based phosphates owing to their rich interesting magnetic properties [11, 12].

We concentrate here on the magnetic properties of $LiNi_2P_3O_{10}$. A close look at the structure of $LiNi_2P_3O_{10}$ (see sec. III) suggests that Ni ($S = 1$) atoms might be placed in a quasi-1D or quasi-2D arrangement. Therefore an investigation of its magnetic properties is interesting. We have therefore carried out magnetic susceptibility and specific heat studies on pure $LiNi_2P_3O_{10}$ and the effect of magnetic ($S = 1/2$) and non magnetic ($S = 0$) impurities on the parent system.

Magnetic susceptibility data are indicative of an antiferromagnetic interaction between $Ni^{2+}$ with exchange constant $J/k_B$ around -5 K. The system orders antiferromagnetically around 7 K as seen from susceptibility and heat capacity data. A change in the slope of the low temperature magnetisation with applied field is suggestive of an approach to a spin-flop transition which is very similar to that observed in a recently investigated $S = 1$ system [13]. The magnetic entropy change at the ordering temperature (as determined from measured heat capacity data) is consistent with the expected Rln3 for an $S = 1$ system ($R$ is the gas constant).

## II. EXPERIMENTAL DETAILS

Polycrystalline samples of $LiNi_{2(1-x)}Zn_{2x}P_3O_{10}$ ($0 \leq x \leq 0.15$) and $LiNi_{2(1-y)}Cu_{2y}P_3O_{10}$ ($0 \leq y \leq 0.05$) were prepared following the conventional solid state reaction route from the starting materials $Li_2CO_3$ (99.999%, Alfa), NiO (99.998%, Alfa) and $(NH_4)_2HPO_4$ (99.9%, Aldrich). The stoichiometric mixture was ground thoroughly before being heated gradually to $300^0C$ for 6 h followed by heating to $500^0C$ for 6h. The mixture was reground, pelletized and fired at $700^0C$ for 36 h followed by regrinding and pelletization and firing at $770^0C$ for 36 h. The final product is found to be yellow in colour.

X-ray diffraction data of the polycrystalline samples were collected on a PANalytical X'Pert pro powder diffractometer equipped with a Cu target ($\lambda$ = 1.54182 Å) coupled with a nickel filter in the θ-2θ mode. The pattern was recorded in a step scanning mode between $7^0$ to $70^0$ with a $0.017^0$ step width and a step time of 40 seconds. The samples investigated here were found to be single phase and structural parameters were refined using Fullprof software [14]. Magnetisation $M$ was measured as a function of applied magnetic field $H$ ($0 \leq H \leq 90$ kOe) and temperature $T$ ($2 \leq T \leq 300$ K) using a vibrating sample magnetometer of Quantum Design Physical Property Measurement System (PPMS).

We have performed specific heat ($C_p$) measurements using a commercial Quantum Design PPMS on thin, flat pellet samples in the temperature range ($1.8 \leq T \leq 100$ K) with various applied magnetic fields. Every sample measurement was preceded by a measurement of the addenda in the same temperature range following a thermal relaxation method. The sample heat capacity is obtained after subtracting the addenda heat capacity from total heat capacity.

## III. RESULTS AND ANALYSIS
### A. Crystal structure and lattice parameters

The $LiNi_2P_3O_{10}$ structure consists of Ni chains bridged by $PO_4$ tetrahedra but some Ni atoms are additionally linked by oxygen (see Fig. 1). The P-O-P angle in the $PO_4$ tetrahedra is 142.5°. The average Ni-O bond distance is 2.07Å and the nearest Ni-Ni bond distance is 3.023 Å.

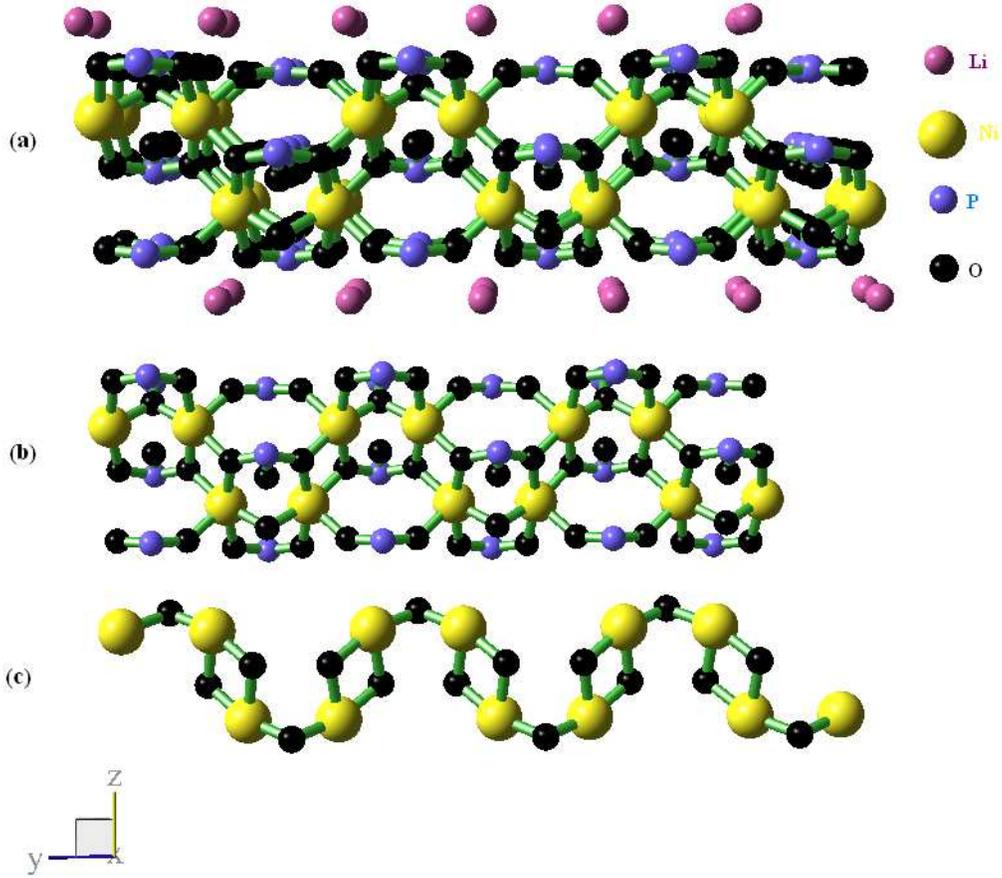

Fig.1. (Colour online) Different views of crystal structure of $LiNi_2P_3O_{10}$.

In the absence of any calculations estimating the relative magnetic exchange interaction between the various Ni atoms, we suggest different ways of looking at the structure in terms of the possible magnetic linkages. One possible way to look at the structure is to think of Ni atoms as forming chains along the crystallographic *b*-direction. As per Goodenough-Kanamori rules, the super exchange interactions are antiferromagnetic in a situation wherein the transfer of electron takes place due to overlap of the localized half filled orbitals of magnetic ions with oxygen orbitals. In $LiNi_2P_3O_{10}$, $Ni^{2+}$ has $e_g^2$ electronic configuration and the angle between nearest Ni-O-N is $101^0$. So one would expect a weak antiferromagnetic interaction arising due to the super exchange interaction between $e_g^2$ and surrounding 2p electrons of $O^{2-}$.

However, there appear to be next-nearest-neighbour and next-next-nearest-neighbour linkages through the PO$_4$ tetrahedra. It is also possible that the Ni atoms which are coupled through PO$_4$ tetrahedra as also through oxygen, form a dimer and then the dimers are coupled. Alternatively, the chains mentioned above are perhaps coupled in the crystallographic *a*-direction in which case the geometry is more like a planar one. Clearly, the case is a complicated one and an unambiguous resolution from experiments may not be possible. On the other hand, it seems reasonably clear that this magnetic system is low-dimensional and this has motivated us to investigate further the properties of the undoped and doped analogs.

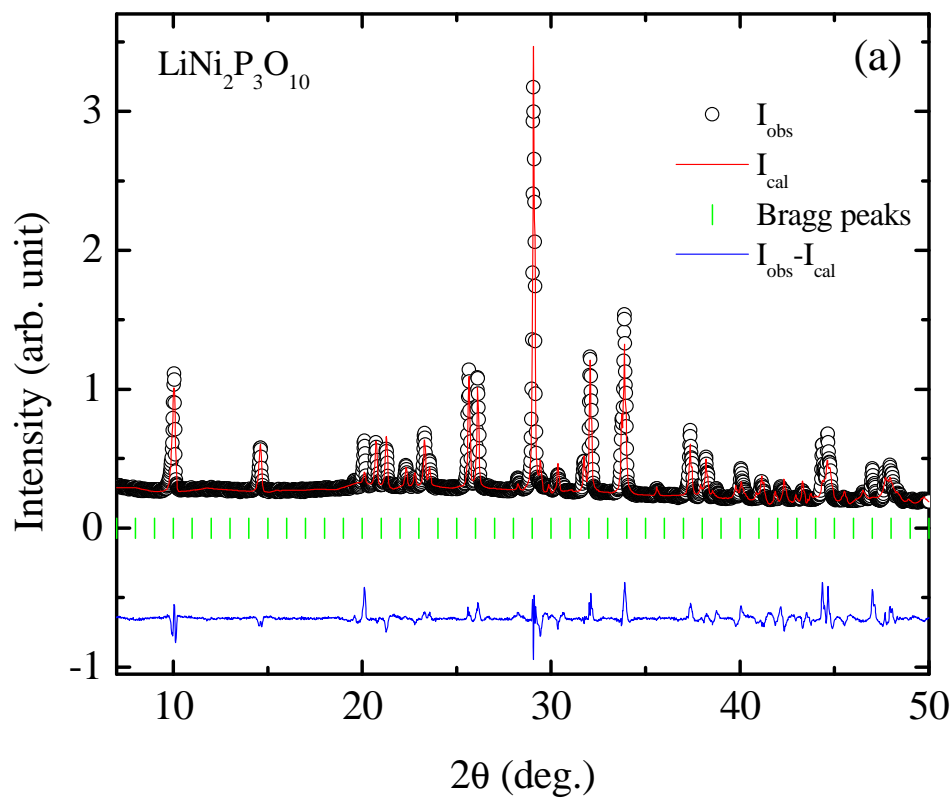

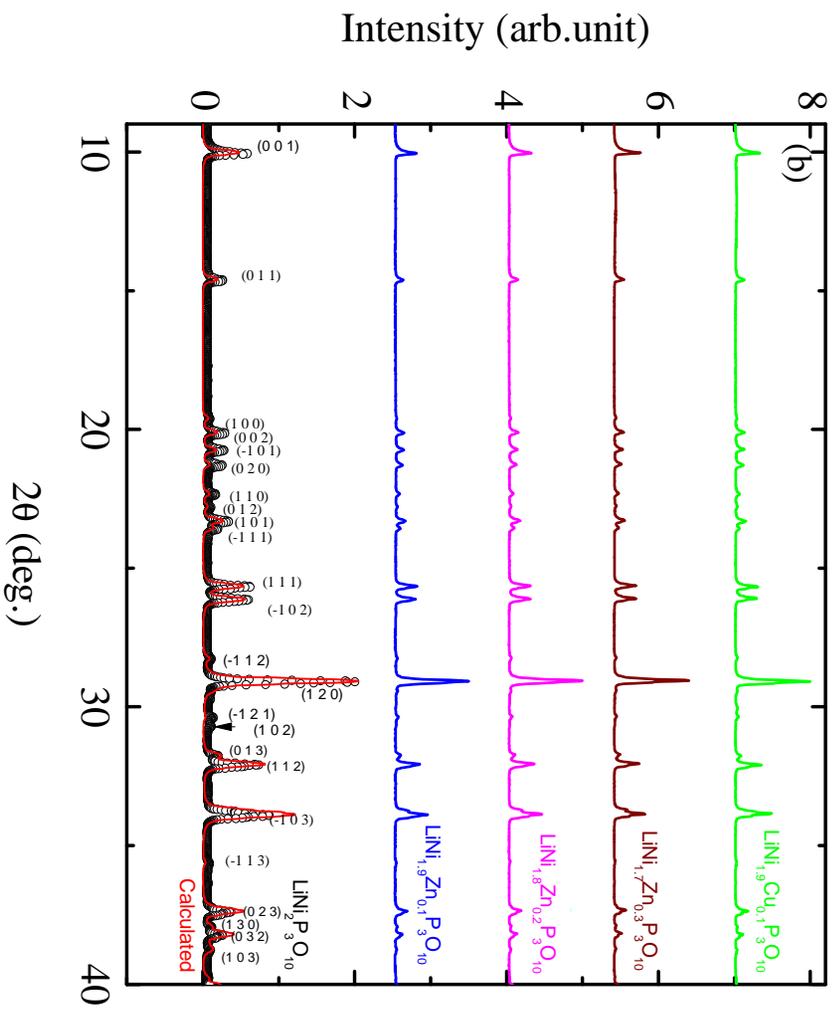

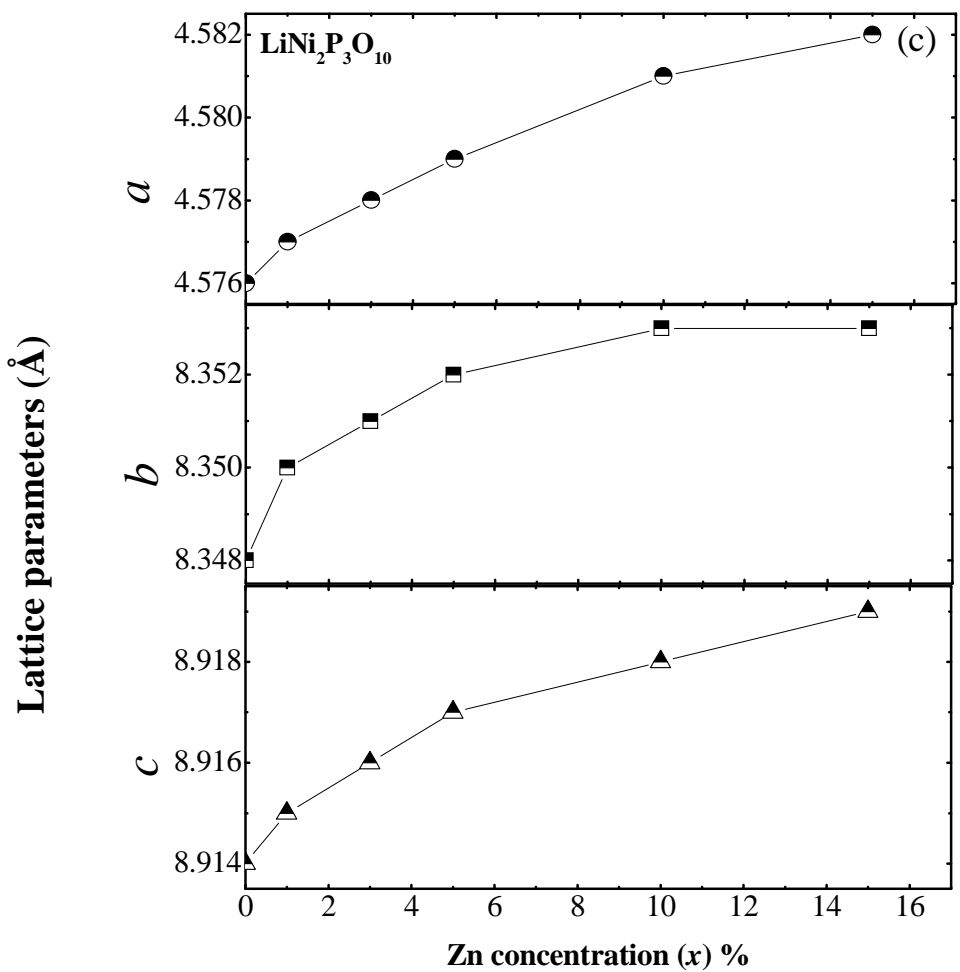

Fig. 2(a).(Colour online) X-ray diffraction spectra of LiNi$_2$P$_3$O$_{10}$ with Rietveld refinement (b) X-ray diffraction pattern for and Zn and Cu doped samples (c) Lattice parameters as a function of Zn doping concentration (%) where lines are drawn as a guide to the eye.

The results of x-ray diffraction on the pure, Zn doped and Cu doped compounds are shown in Fig. 2. No impurity peaks were observed for Zn-doping $x \leq 0.15$ and Cu-doping $y \leq 0.05$. The peaks for all the samples are indexed on the basis of a monoclinic space group P2$_1$/m with room temperature lattice parameters $a = 4.576$ Å, $b = 8.351$ Å, $c = 8.918$ Å obtained using Fullprof Rietveld refinement program [14]. The obtained refinement parameters are $R_p$=16 and $R_{wp}$=12. The lattice parameters for the undoped sample are consistent with the previously reported data [15, 16]. The lattice parameters increase steadily with Zn-doping (as shown in Fig. 2(c)). The lattice parameters for the 5 % Cu doped sample are similar to those of the 5% Zn doped sample.

## B. Magnetic susceptibility and specific heat

Undoped $LiNi_2P_3O_{10}$

### (i) Magnetic susceptibility

The temperature $T$ dependence of the magnetisation $M$ in applied magnetic fields $H = 5$ kOe and 80 kOe was measured for $LiNi_2P_3O_{10}$ and the resulting magnetic susceptibility $\chi(T) = M/H$ is shown in Fig. 3(a). These data show a broad maximum at around 10 K. In the case of AF order the anomaly in susceptibility is sharp unlike the broad one seen by us. This is indicative of short range order. At first sight, there does not seem to be any other anomaly but a closer investigation, especially plotting $d\chi/dT$ as in the inset of Fig. 3(b) shows a sharp maximum at around 7 K. This is suggestive of some kind of magnetic long range ordering. It is also seen that the broad maximum shifts towards low temperature with fields. We now attempt to analyse the $\chi(T)$ data quantitatively. The temperature dependence of $\chi$ is fitted with the Curie-Weiss law $\chi(T) = \chi_0 + C/(T-\theta)$ in the high temperature (200-300K) range which gives $C = 1.4$ cm$^3$ K/mole (which is somewhat larger that the value of 1 expected for $S = 1$ moments), $\theta = -3.6$ K, $\chi_0 = -3 \times 10^{-5}$ cm$^3$/mole. The low and negative value of $\theta$ is indicative of weak antiferromagnetic interaction between $Ni^{2+}$ spins in $LiNi_2P_3O_{10}$.

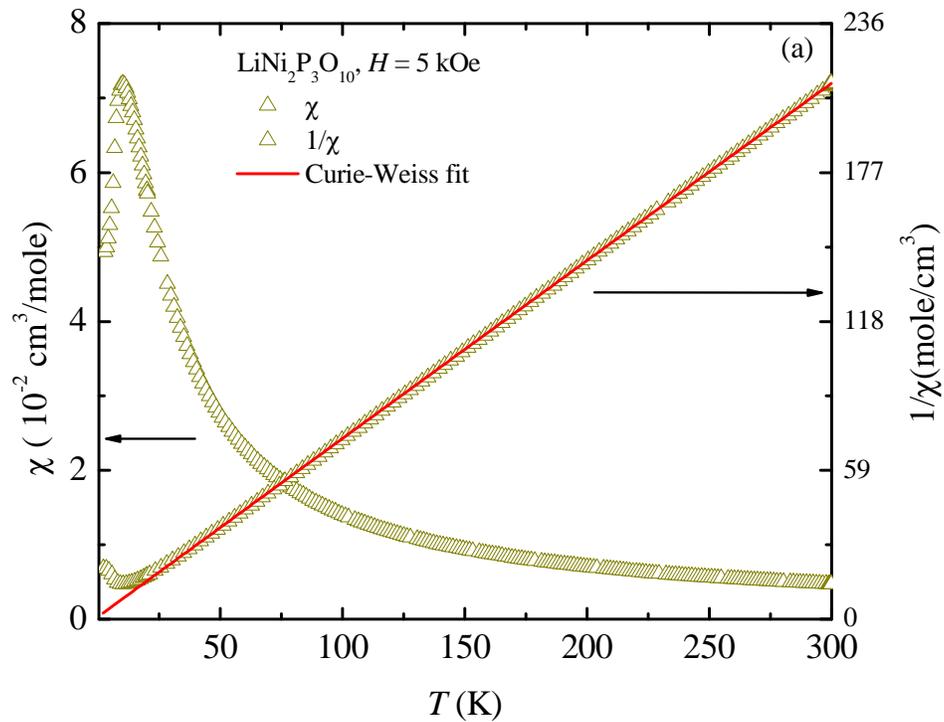
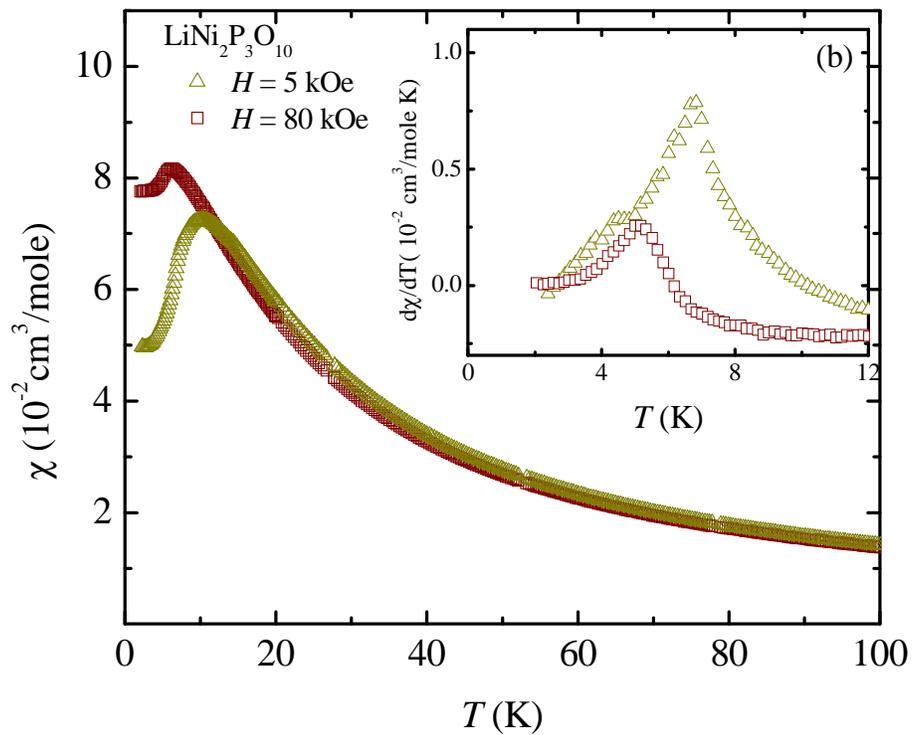

Fig. 3(a). (Colour online) Temperature dependence of magnetic susceptibility of LiNi$_2$P$_3$O$_{10}$ with the Curie-Weiss fit discussed in the text (b)Temperature dependence of magnetic susceptibility in LiNi$_2$P$_3$O$_{10}$ in 5 kOe (open triangles) and 80 kOe (open circles). The inset shows d$\chi$/d$T$ as a function of temperature in the low temperature range.

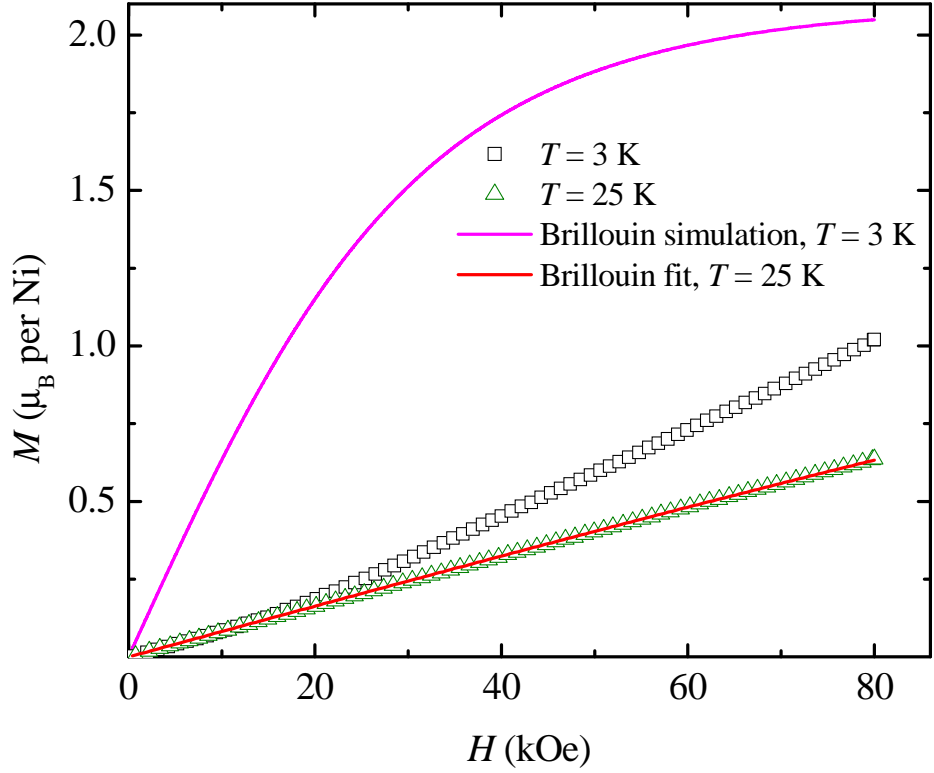

Fig. 4. (Colour online) Magnetisation isotherms in LiNi$_2$P$_3$O$_{10}$ at 3 K and 25 K with the fits as discussed in the text.

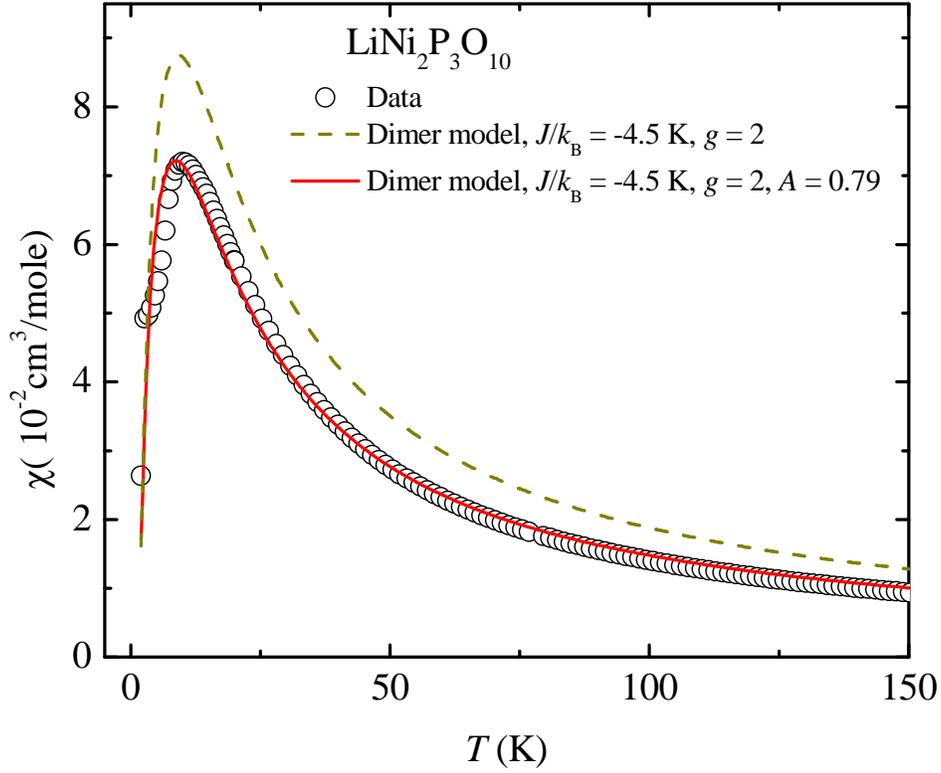

Fig. 5. (Colour online) Magnetic susceptibility in LiNi$_2$P$_3$O$_{10}$ with a fit to the dimer model. Only some of the experimental data points are displayed for better visual clarity.

Likewise the effective moment $\mu_{\text{eff}} = \sqrt{\dfrac{3k_B C}{N_A}} = 3.34~\mu_B$, where $C$ is the Curie-constant, $k_B$ is Boltzmann constant and $N_A$ is the Avogadro number. This is higher than the expected value $\mu_{\text{eff}} = 2.83~\mu_B$ for $S = 1$, $g = 2$ in the case of Ni$^{2+}$. On increasing $H$ from 5 kOe to 80 kOe, the location of the broad maximum is reduced from 10 K to 6.1 K and the anomaly in $d\chi/dT$ appears at 5.1 K instead of 6.8 K in the undoped system as shown in Fig. 3(b). The decrease with field of the temperature at which there is a dχ/d*T* anomaly indicates the presence of an antiferromagnetic (AFM) transition.

The field dependence of the magnetisation isotherm is linear at high temperature as expected for a paramagnet. At temperatures below the maximum, the magnetisation is much lower than that expected for a paramagnet (see the Brillouin function in Fig. 4) which indicates the presence of an AFM state. Further, for the 3 K data, the increase in slope with field might be an indication of spin–flop transition. We note that it is difficult to get correct information regarding a spin-flop transition from polycrystalline samples. We are tempted to make the guess in our case because similar behaviour has been found in other polycrystalline samples [13].

Concerning the analysis of the *T*-dependence of the susceptibility, the Curie-Weiss fit works well in the high-*T* range but obviously cannot reproduce the maximum at 10 K. On the other hand, for a low-dimensional material, a broad maximum is expected. As we have speculated in the section on crystal structure, a dimer model is one of the possibilities for modeling the magnetic properties of this compound.

For a system of isolated $S = 1$ dimers, governed by the Heisenberg interaction $H = -2J\, S_1 \cdot S_2$ (where $S_1$ and $S_2$ are the spins comprising the dimer and *J* is the exchange interaction), the temperature dependence of magnetic susceptibility can be expressed as

$$\chi(T) = \frac{N_A g^2 \mu_B^2}{3 k_B T} \frac{\sum_{S=0}^{2} S(S+1)(2S+1)\exp\left(-\frac{E}{k_B T}\right)}{\sum_{S=0}^{2} (2S+1)\exp\left(-\frac{E}{k_B T}\right)} \quad (1)$$

Taking energy spectrum of the spin states $E = -JS(S+1)$ with $S = S_1 + S_2$, we obtain

$$\chi(T) = \frac{N_A g^2 \mu_B^2}{3 k_B T} \frac{\left[6\exp\left(\frac{2J}{k_B T}\right) + 30\exp\left(\frac{6J}{k_B T}\right)\right]}{1 + 3\exp\left(\frac{2J}{k_B T}\right) + 5\exp\left(\frac{6J}{k_B T}\right)} \quad (2)$$

Whereas a fit to the above dimer model is not so good, we obtained a better fit with a prefactor $A$ (which might account for deviations from the ideal dimer model) in the above equation giving the fitting parameters $J/k_B$ = -4.5 K, and A = 0.79 (fixing $g$ = 2).

(ii) Heat Capacity

The temperature dependence of heat capacity for $LiNi_2P_3O_{10}$ is shown in Fig. 6. The temperature (~7 K) at which a λ-like anomaly is observed in the $C_p$ ($T$) data is the same at which there is an anomaly in the $d\chi(T)/dT$ data. This is an evidence of long-range AFM order.

We would now like to extract the magnetic contribution to the measured specific heat. In situations where a non-magnetic analog is available, the lattice contribution can be deduced from it. Due to non-availability of a non-magnetic analog for $LiNi_2P_3O_{10}$, we have fitted the experimental specific heat data with a combination of four Debye functions [17] in the high temperature range (60-100 K) where the heat capacity is dominated by lattice contributions and then extrapolated to 2 K as shown in Fig. 6 (a).

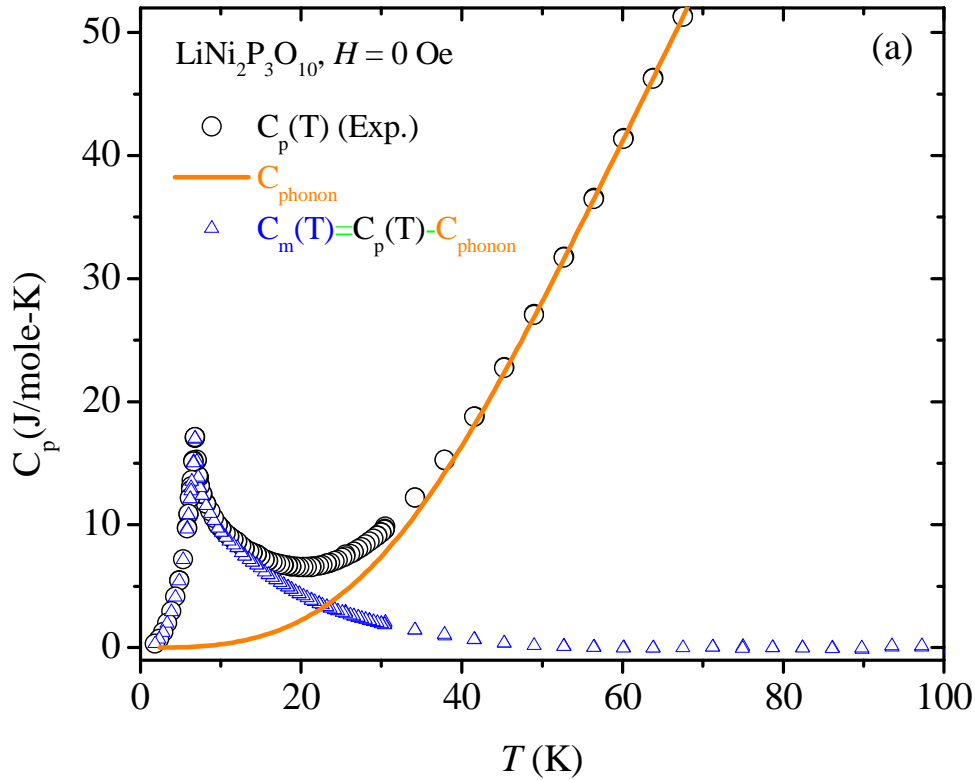

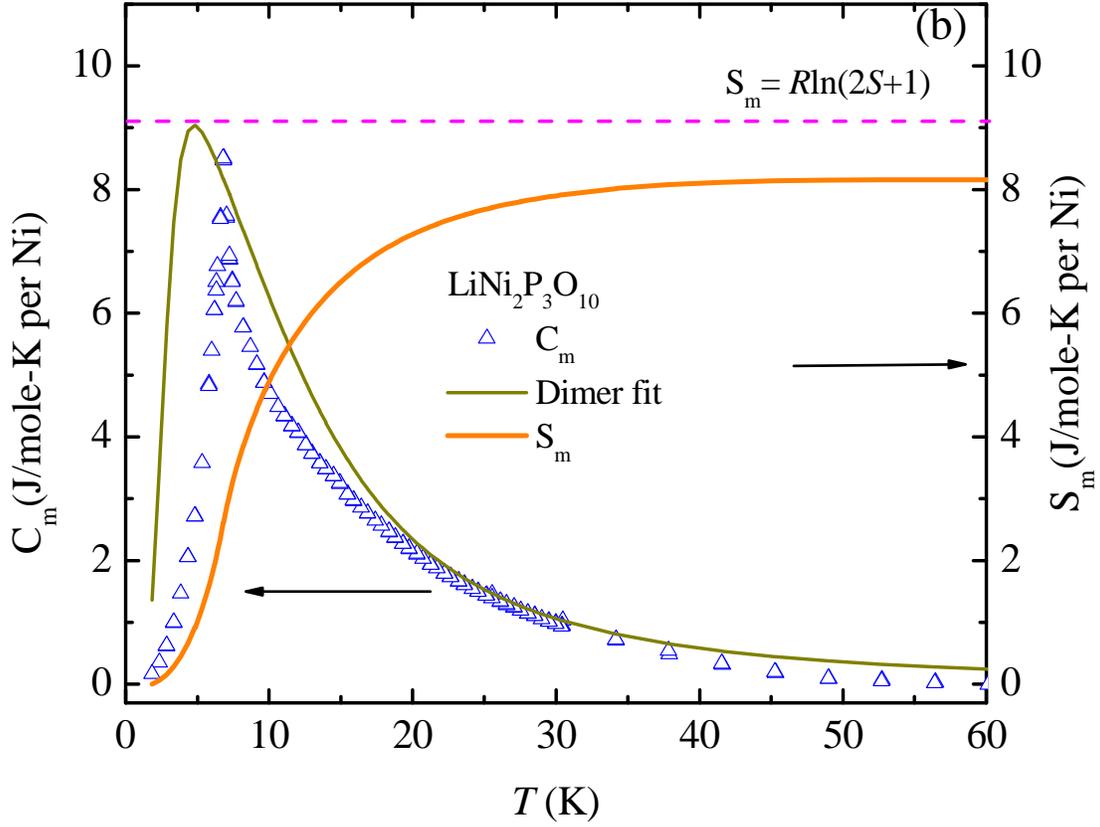

Fig. 6(a). (Colour online) Specific heat in LiNi$_2$P$_3$O$_{10}$ with a fit to Eq. (3) (b) Magnetic specific heat with a fit to the dimer model as explained in the text. The magnetic entropy is also shown (see right y-axis)

The fitting function used is

$$C_p(T) = 9R \sum_{i=1}^{4} C_i \left(\frac{T}{\theta_D^i}\right)^3 \int_0^{x_D^i} \frac{x^4 e^x}{(e^x-1)^2} dx \qquad (3)$$

Where $\theta_D^i$ is Debye temperature, $x_D^i = \theta_D^i/T$, The fitting parameters are $C_1 = 0.45\pm0.02$, $\theta_D^1 = 339\pm3$ K, $C_2 = 0.5\pm0.03$, $\theta_D^2 = 352\pm2$ K, $C_3 = 0.34\pm0.02$, $\theta_D^3 = 449\pm5$ K, $C_4 = 0.27\pm0.05$, $\theta_D^4 = 345\pm4$ K, $R = 8.314$ J/mole-K is the gas constant. The sum of Debye functions accounts for the lattice contribution to specific heat with different Debye temperatures due to different elements forming $LiNi_2P_3O_{10}$. Magnetic contribution to the heat capacity is obtained by subtracting the lattice heat capacity from the total heat capacity.

The magnetic heat capacity $C_m(T)$ thus obtained represents the intrinsic magnetic contribution of $LiNi_2P_3O_{10}$ and is shown in Fig. 6 (b). Magnetic specific heat is fitted with the dimer model as given below.

$$C_m(T) = 12R\left(\frac{J}{k_BT}\right)^2 \frac{\exp\left(\frac{2J}{k_BT}\right)+15\exp\left(\frac{6J}{k_BT}\right)}{1+3\exp\left(\frac{2J}{k_BT}\right)+5\exp\left(\frac{6J}{k_BT}\right)} \qquad (4)$$

Taking R the universal gas constant to be 8.314 J/mole-K, we get $J/k_B = -5.5$ K. The fit is poor and indicates inapplicability of the simple isolated dimer model.

A plot of the temperature dependence of magnetic entropy $S_m(T) = \int \frac{C_m(T)}{T}dT$ is given in Fig.6 (b). Rise of $S_m(T)$ with temperature is associated with a transition from ordered state to a disordered state. The associated entropy change is expected to be $R\ln(2S+1) = 9.13$ J/mole-K where $S$ is the spin of the magnetic ions. Considering the uncertainty involved in determining the lattice specific heat, the value obtained by us (8.1 J/mole-K) is in reasonable agreement with the expected value.

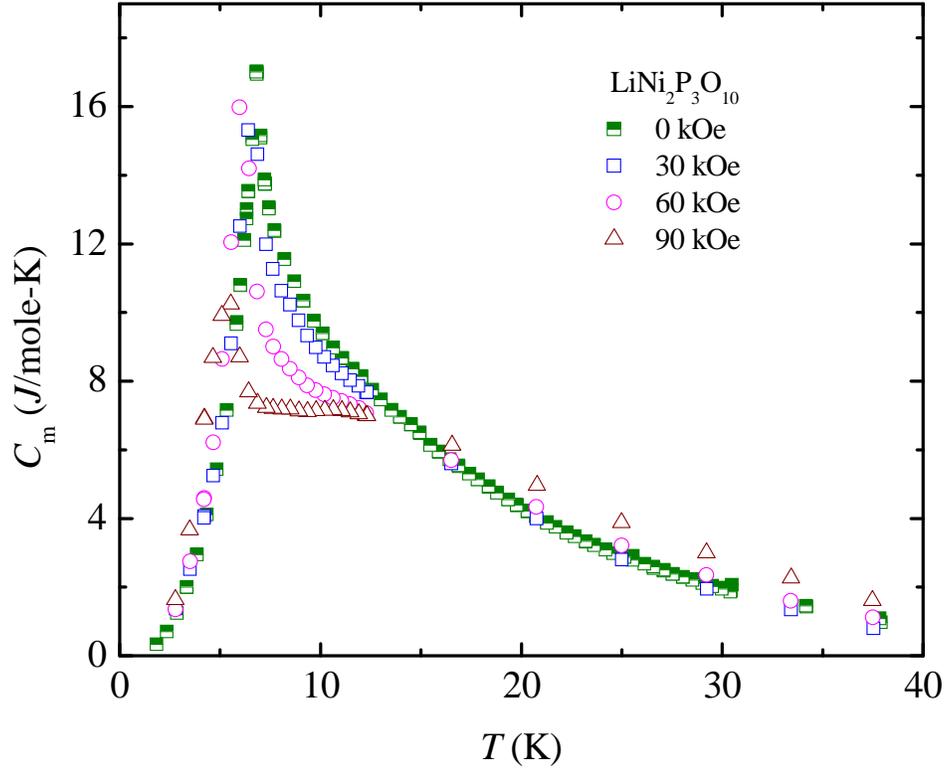

Fig. 7. (Colour online) Temperature dependence of magnetic specific heat is shown for $LiNi_2P_3O_{10}$ in various applied fields.

The field dependence of specific heat is shown in Fig. 7. The ordering temperature $T_N$ is found to be suppressed with the field. Also, there is a significant suppression of the magnetic specific heat just above $T_N$ with the application of magnetic field. This perhaps indicates the suppression of the short-range order in a field.

# Zn and Cu doped $LiNi_2P_3O_{10}$

(i) Magnetic Susceptibility

Let us now consider the effects of impurity substitution on the magnetic properties of $LiNi_2P_3O_{10}$. The magnetic susceptibility data in Zn doped $LiNi_2P_3O_{10}$ are shown in Fig. 8 (a). The broad maximum in $\chi(T)$ shifts to lower temperatures and the ordering temperature reduces with increasing doping [as shown in Fig. 8 (b)]. The change of the slope of magnetisation with applied magnetic field is indicative of an approach to a spin-flop type transition. Surprisingly, no Curie-like upturn appears at low-$T$ with increasing Zn-doping which is expected in dimers or low dimensional quantum spin systems.

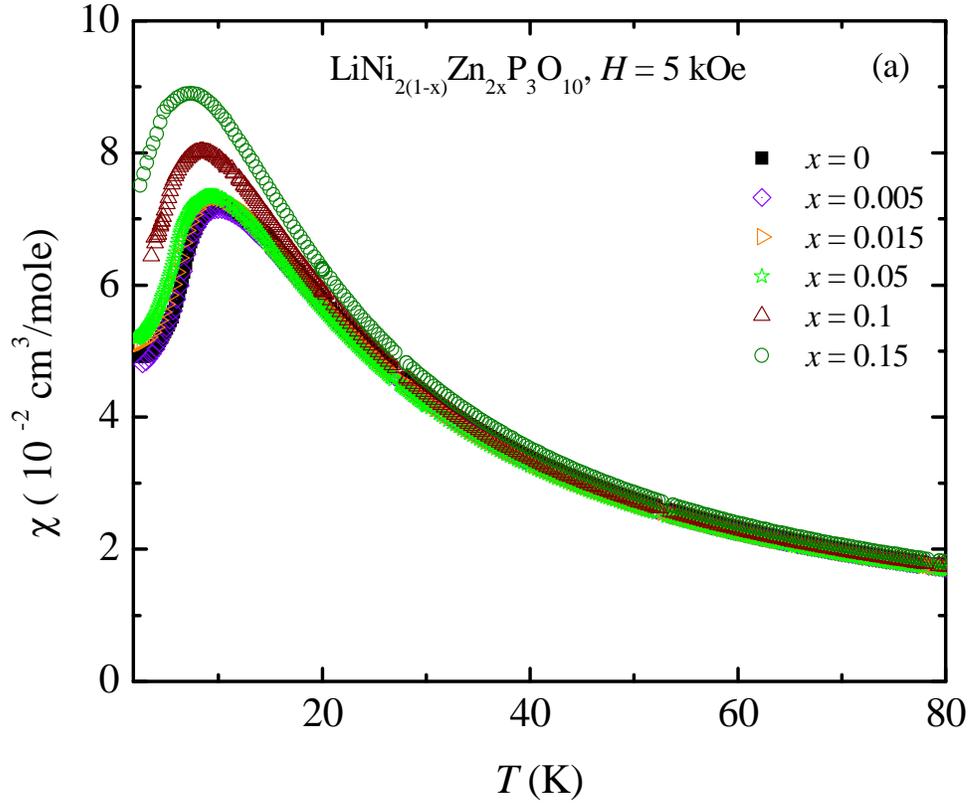

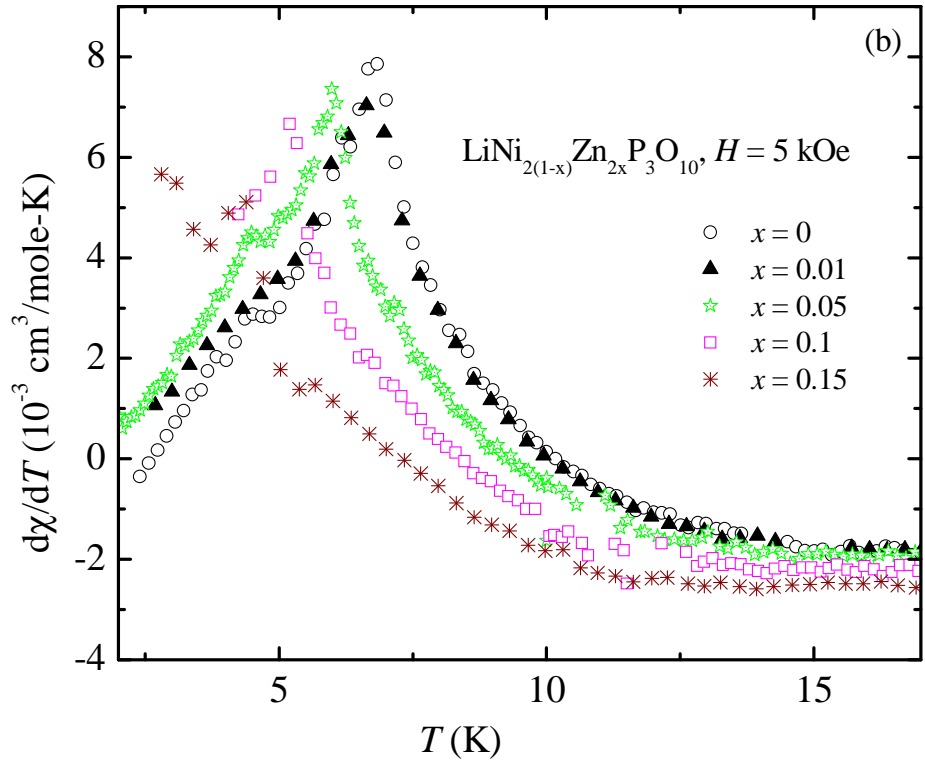

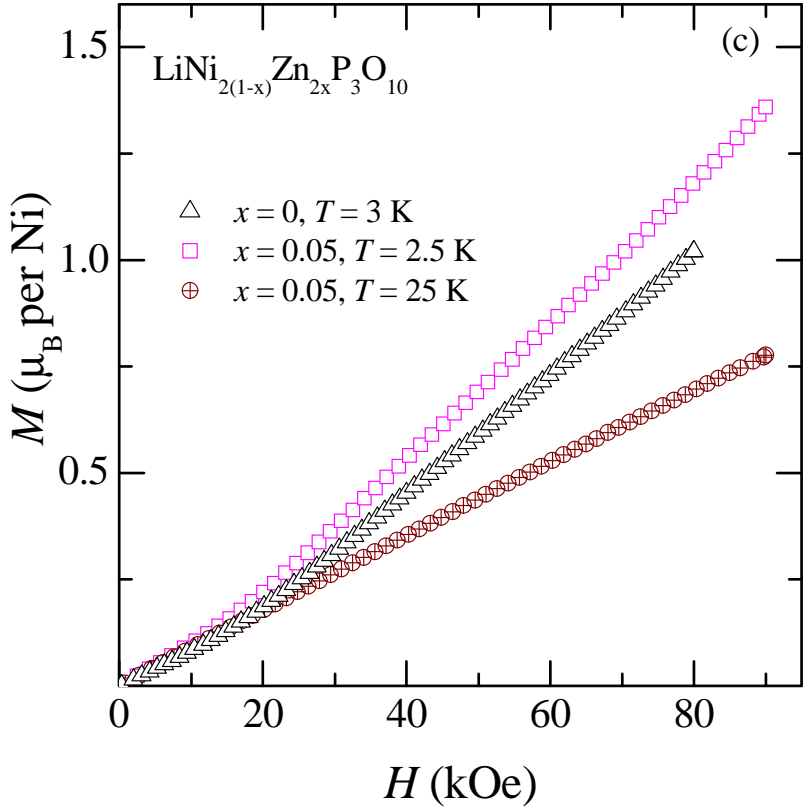

Fig. 8(a). (Colour online) Magnetic susceptibilities in Zn doped samples (b) d$\chi$/d$T$ in Zn doped samples (c) Magnetisation isotherm at various temperatures in LiNi$_{1.9}$Zn$_{0.1}$P$_3$O$_{10}$.

In quantum antiferromagnets, introduction of $S = 0$ vacancies result in induced local moments on the neighbours and for a sufficient concentration this can give rise to spin-freezing (order from disorder) [18, 19]. Theoretically, Sandvik *et al*., have shown that impurity susceptibility diverges at certain points in the wave vector space [20]. Various disorder driven phase transitions have been observed in doped 2D antiferromagnets [21] and order-disorder transition in these systems is driven by site dilution [22]. In the prototypical ladder SrCu$_2$O$_3$, Zn impurities give rise to a cluster AF state [23]. Also, in the coupled two-leg ladder BiCu$_2$PO$_6$, Zn and Ni substitutions give rise to spin-freezing at low-$T$ [24].

In contrast, the present system LiNi$_2$P$_3$O$_{10}$ behaves like a classical 3D antiferromagnet. Here, the introduction of Zn does not give rise to a low-$T$ Curie term (no induced magnetism). In fact, the $T_N$ decreases with increasing substitution unlike the case in low-dimensional quantum AF systems where disorder (from substituents) induced order. This further indicates that magnetic interactions are of a 3D nature in this system. A Similar explanation is also valid in the case of Cu$^{2+}$ ($S = 1/2$) doping discussed below.

The magnetic susceptibility data in the Cu doped are shown in Fig. 9. The substitution of Cu$^{2+}$ ($S = 1/2$) in place of $S = 1$ does not affect much the parent system. The variation of broad maximum and ordering temperature is very similar to Zn doped samples.

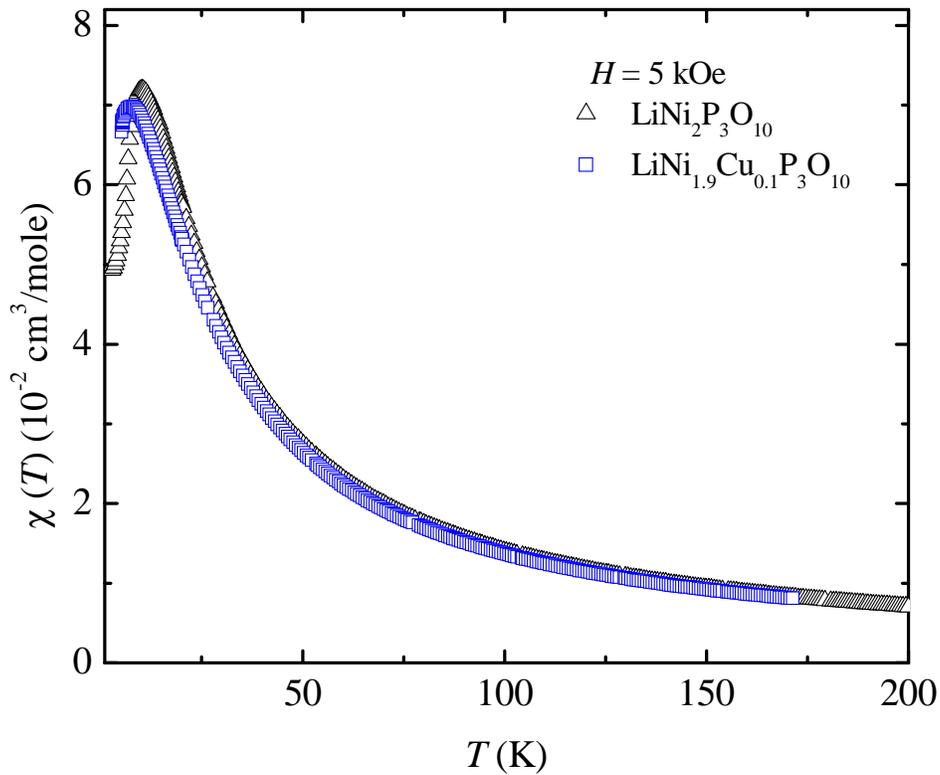

Fig. 9. (Colour online) Magnetic susceptibility in Cu doped LiNi$_2$P$_3$O$_{10}$.

### (ii) Heat Capacity

Specific heat data for different Zn dopings were taken in zero field and in a magnetic field ($0 \leq H \leq 90$ kOe) for Zn % doped samples. Specific heat data also shows a λ-like anomaly and broad maximum in Zn doped samples at the corresponding temperatures as observed in the magnetic susceptibility data.

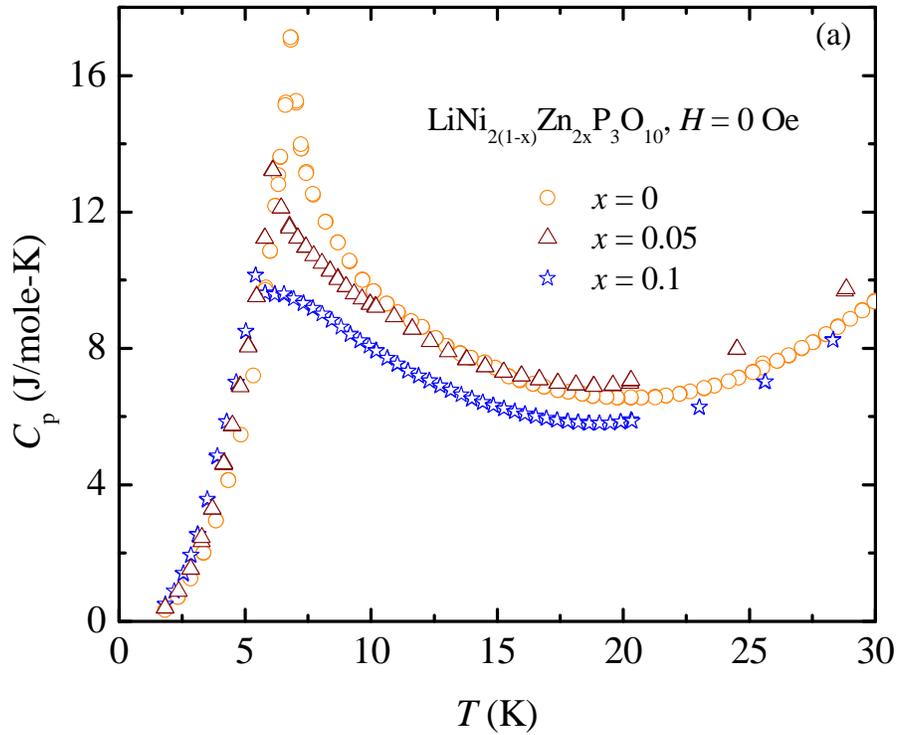

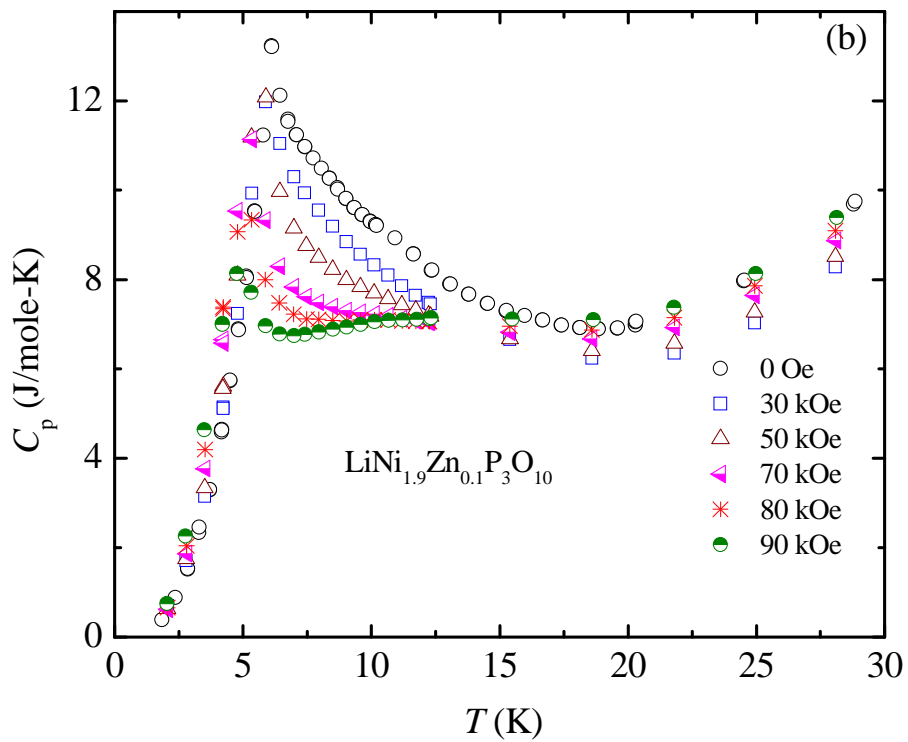

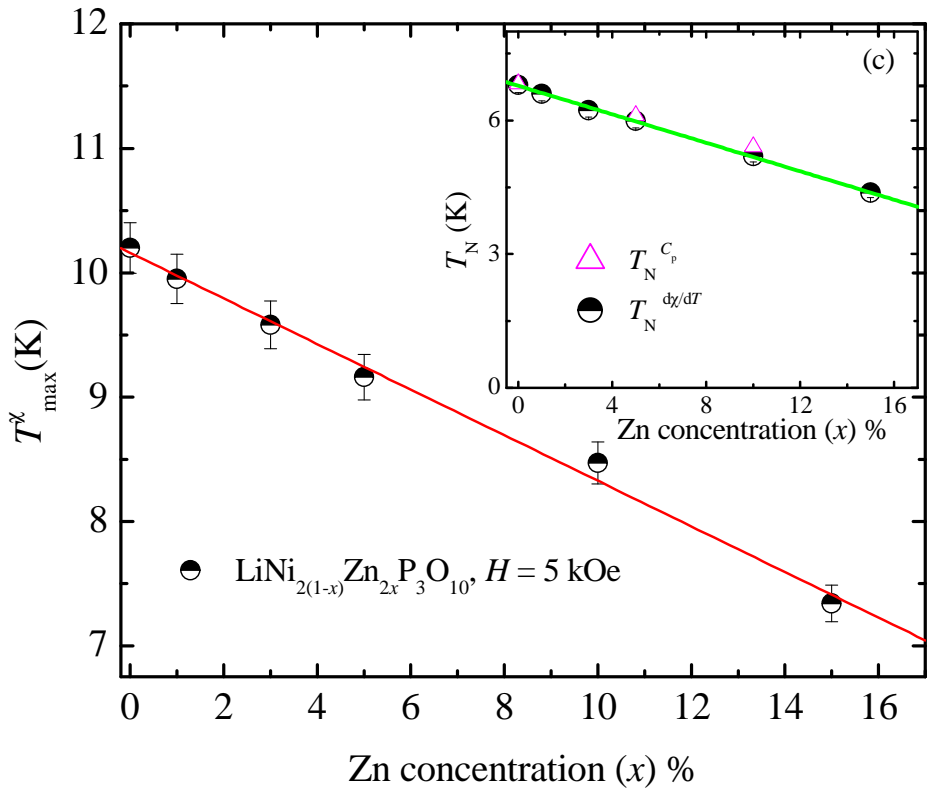

Fig. 10 (a). (Colour online) Temperature dependence of specific heat of Zn doped samples in zero field (b) Specific heat in Zn 5% sample in various applied fields are shown (c) Position of broad maximum [$T^\chi_{max}$ (K)] in magnetic susceptibility data vs Zn concentration (%) and the inset shows ordering temperature (obtained from $d\chi/dT$ and $C_p(T)$ at zero field) vs Zn concentration (%) and the solid lines are guides to eye.

Whereas there is a reduction in $T_N$ on Zn substitutions, the overall nature of the $T$-variation of $C_p$ remains the same as in undoped LiNi$_2$P$_3$O$_{10}$. The ordering temperature reduces with doping and magnetic field. The nearly linear decrease of $T_N$ is ascribed to a dilution of the AFM lattice by non-magnetic Zn substitution. Heat capacity data in Cu 5% doped samples were taken in the magnetic field range ($0 \leq H \leq 90$ kOe ). Like in the case of undoped and Zn doped samples, the ordering temperature in this case also decreases with field. Interestingly, in all cases (pure, Zn-doped or Cu-doped) the application of a magnetic field suppresses the heat capacity above $T_N$ and the anomaly at $T_N$ seems to be sharper. This is perhaps related to the movement of $T^\chi_{max}$ (position of the broad maximum and hence onset of short-range-order) to lower values with a reduction of the crossover region. The low temperature specific heat (below $T_N$) follows $T^3$ behaviour which is expected on the basis of spin wave theory [25]. The field dependent specific heat results enable us to draw a $H$-$T$ phase diagram which is shown Fig. 12.

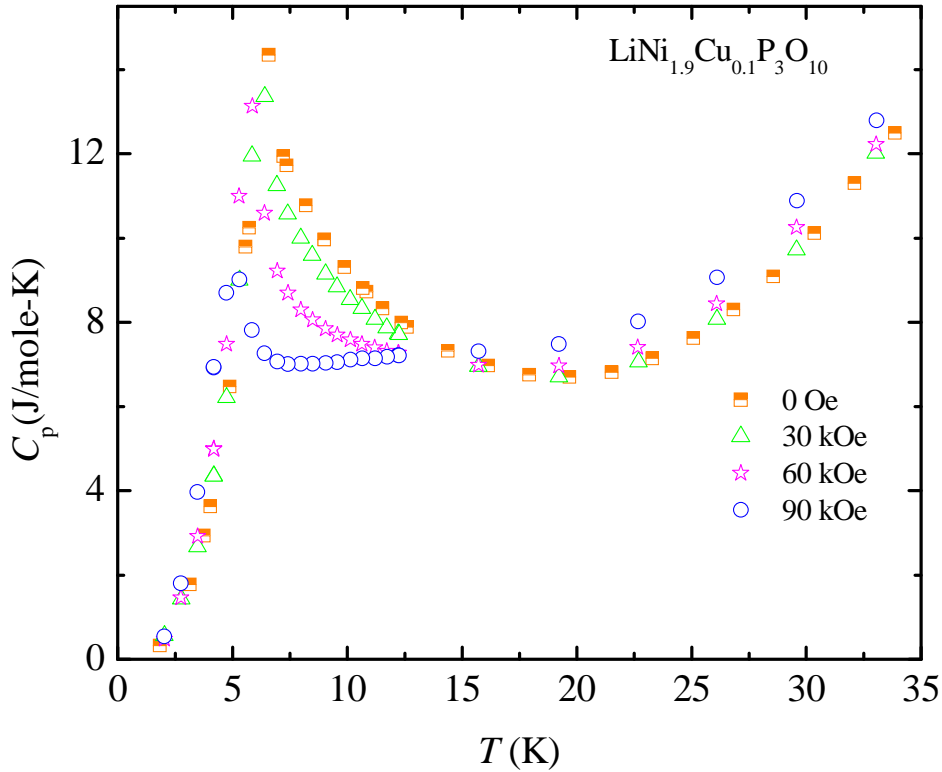

Fig. 11. (Colour online) Temperature dependence of specific heat in 5% Cu doped $LiNi_2P_3O_{10}$ in various applied magnetic fields.

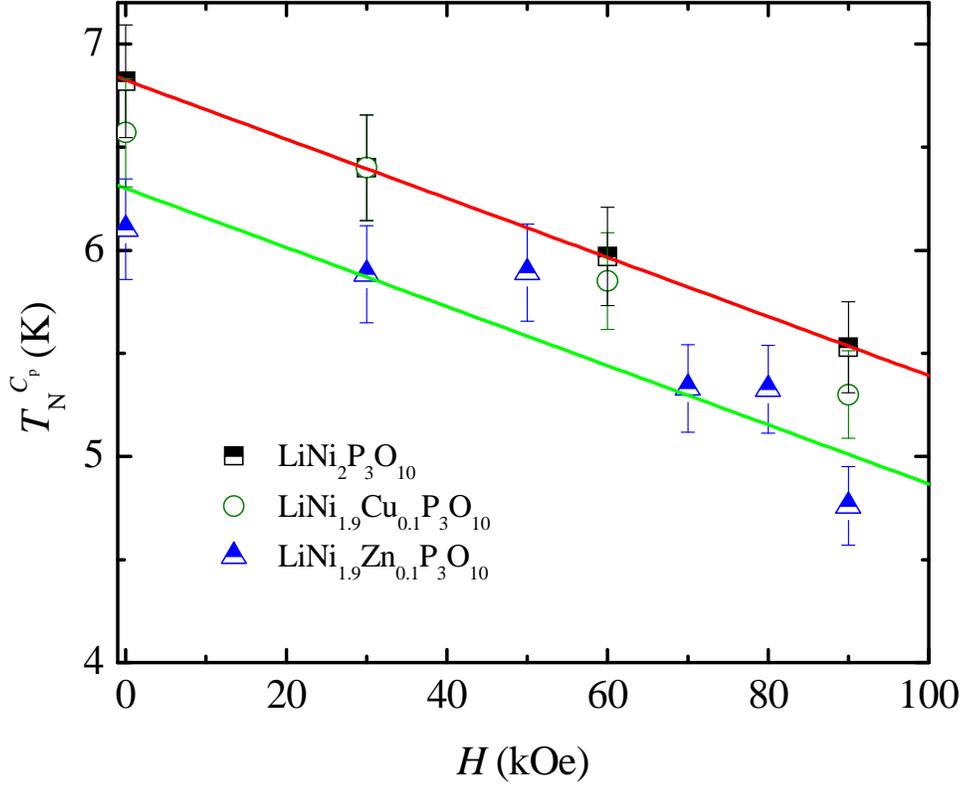

Fig. 12. The *H-T* phase diagram drawn from the field dependence of specific heat for undoped, Zn-doped and Cu-doped samples. The solid lines are drawn as a guide to the eye.

## IV. CONCLUSION

In conclusion, we have presented results of our measurements of the temperature dependence of magnetic susceptibility and heat capacity in single phase samples of undoped, Zn-doped and Cu-doped samples of $LiNi_2P_3O_{10}$.

The parent system shows a broad maximum in magnetic susceptibility at around 10 K suggestive of low dimensionality of the system. But the $\lambda$-like anomaly at about 7 K observed in the specific heat and $d\chi/dT$ data are indicative of a cross over to 3D behaviour and long range ordering at $T_N$ ~ 7 K which might be due to weak interchain/interdimer interaction. The change in slope of the low temperature magnetisation data might be associated with some kind of spin-flop transition. The change in magnetic entropy at $T_N$ is consistent with the expected $R\ln 3$. The fit of magnetic susceptibility data to the dimer model yields a coupling constant $J/k_B$ ~ -5 K due to antiferromagnetic interactions between $Ni^{2+}$. Non-magnetic impurity $Zn^{2+}$ ($S = 0$) and magnetic impurity $Cu^{2+}$ ($S = 1/2$) reduce the ordering temperature which can be due to a dilution effect. We did not find any extra Curie term in the low-$T$ magnetic susceptibility and the behaviour of the low-temperature magnetisation is very similar to that of the undoped system. The reduction of $T_N$ with applied magnetic field in the doped systems is very similar to the parent system. Theoretically, band structure (LDA+U) calculations may provide information about the relative importance of the various exchange paths leading the LRO for the system presented here. We finally note that systems with a low-dimensionality which show a crossover to 3D are rich in physics and parameters (such as substitutions, pressure, field, etc.) which can move the crossover temperature would be interesting to investigate in the future to understand the complexities of magnetic linkages in real materials.

# References

1. Bednorz J G and Muller K A 1986 Z. *Phys*. B **64** 189.
2. Anderson P W 1987 *Science* **235** 1196.
3. Dagotto E and Rice T M 1996 *Science* **271** 618.
4. Sachdev S 2000 *Science* **288** 475.
5. Bethe H A 1931 Z. *Phys*. **71** 205.
6. Mikeska H J, Ghosh A, and Kolezhuk A K 2004 *Phys. Rev. Lett*. **93** 217204.
7. Birgeneau R J, Als-Nielsen J, and Shirane G 1977 *Phys. Rev.* B **16** 280.
8. de Jongh L J, 1985 *Solid State Commun.* **53** 731.

9. Neel L 1936 *Ann. Phys*. **5** 232 and Neel L 1952 Proc. *Phys. Soc., London, Sect*. A **65** 869.
10. Poulis N J and Hardeman G E G 1952 *Physica (Utrecht)* **18** 315.
11. Nath R, Mahajan A V, Büttgen N, Kegler C, Loidl A and Bobroff J 2005 *Phys. Rev*. B **71** 174436.
12. Nath R, Tsirlin A A, Kaul E E, Baenitz M, Büttgen N, Geibel C, and Rosner H 2008 *Phys. Rev*. B **78** 024418.
13. He Zhangzhen, Chen S C, Lue C S, Cheng Wendan, and Ueda Yutaka 2008 *Phys. Rev. B* **78** 212410.
14. [http://www.ill.fr/dif/Soft/fp/index.html](http://www.ill.fr/dif/Soft/fp/index.html).
15. Erragh F, Boukhari A and Holt E M 1996 *Acta Cryst*.C **52** 1867.
16. Zoubir M, Erragh F, Boukhari A, and A. El Hajbi A 2004 *Phys. Chem. News* **16** 121.
17. Gopal E S R 1966 *Specific heats at low temperatures* (New York: Plenum Press).
18. O. P. Vajk, P. K. Mang, M. Greven, P. M. Gehring, and J. W. Lynn 2002 *Science* **295** 1691.
19. S. W. Cheong, A. S. Cooper, L. W. Rupp, B. Batlogg, J. D. Thompson, and Z. Fisk, 1991 *Phys. Rev. B* **44**, 9739.
20. Sandvik A W, Scalapino D J 1993 *Phys. Rev. B* **47**, 10090.
21. Sandvik A W, 2006 *Phys. Rev. Lett*. **96**, 207201.
22. Sandvik A W, 2002 *Phys. Rev. B* 66, 24418.
23. Azuma M, Fujishiro Y, Takano M, Nohara M, and Takagi H 1997 *Phys. Rev*. B **55** R8658.
24. Koteswararao B, Mahajan A V, Alexander L K, and Bobroff J 2010 *J. Phys: Condens. Matter* **22** 035601.
25. Kubo R 1952 *Phys. Rev*. **87** 568.